# On the origin of heating-induced softening and enthalpic reinforcement in elastomeric nanocomposites

Pierre Kawak,[1] Harshad Bhapkar,[2] David S. Simmons[3,*]

Department of Chemical, Biological, and Materials Engineering, University of South Florida, Tampa, Florida, USA

**ABSTRACT.** Despite a century of use, the mechanism of nanoparticle-driven mechanical reinforcement of elastomers is unresolved. A major hypothesis attributes it to glassy interparticle bridges, supported by an observed inversion of the variation of the modulus $E(T)$ on heating – from entropic stiffening in elastomers to enthalpic softening in nanocomposites. Here, molecular simulations reveal that elastomer enthalpic softening can instead emerge from a competition over preferred nonequilibrium volumes between elastomer and nanoparticulate networks. A theory for this competition accounting for softening of the bulk modulus on heating predicts the simulated $E(T)$ inversion, suggesting that reinforcement is driven by a volume-competition mechanism unique to co-continuous systems of soft and rigid networks.

Mechanical reinforcement of elastomers by nanoparticles is a cornerstone of technological use of elastomeric materials. For over a century, nanoparticulate filler additives – mainly carbon black – have been introduced to elastomers to enhance Young's modulus and other mechanical properties, such as toughness [1]. This has been essential to enabling the use of elastomers in applications from automobile tires to paints and coatings. Despite this importance, the precise mechanism of nanoparticulate reinforcement of elastomers remains unresolved [1,2]. While earlier work aimed to explain these effects via extensions of dilute mixing theory [3–5], over the last several decades the field has concluded that intrinsically non-dilute mechanical percolation of nanoparticles is central to reinforcement [1,6–12]. However, the precise *nature* of this percolation, and the mechanism by which it leads to reinforcement, has remained a matter of debate. Proposed scenarios include crosslinking of particles by polymer chains [7,13–16], direct contact jamming between fillers [17–20], and the formation of glassy interparticle bridges that essentially glue a particulate network together [9,21–27]. The extent to which one of these mechanisms may dominate reinforcement is an enduring question in the physics and design of elastomeric nanocomposites.

Within this literature, one key observation has been viewed as providing especially compelling support for the glassy bridge hypothesis. Neat (unfilled) elastomers stiffen on heating – a signature of the predominantly entropic origin of polymer network elasticity [28]). In contrast, highly nano-reinforced elastomers *soften* on heating [13,29,30], signaling a shift to an enthalpically-dominated mode of elasticity. This shift of thermodynamic origin of the modulus has been treated as strong evidence for the glassy bridge model of reinforcement [22,23,29,31,32], or at least for strong enthalpic attractive interactions between particles [13]. Within the glassy bridge model, reinforcement emerges because of regions of enhanced glass transition temperature ($T_g$) that are proposed to exist in thin domains between nanoparticles. These domains act as 'glue' holding the nanoparticles together [10]. Because glassy elasticity is enthalpic in nature, these bridges soften on heating, leading to a prediction of an inverted temperature dependence of the overall modulus [23]. Given that particle interactions, in the absence of strong attractions or glassy bridges, are essentially athermal, some attractive bridging interaction of this kind is commonly viewed as *necessarily* implied by the inverted temperature dependence of highly reinforced elastomeric nanocomposites. If so, this would suggest that glassy bridging is the central mechanism of nanocomposite reinforcement.

Despite this evidence and several apparent successes of the glassy bridge model [27,33], there remains significant controversy regarding the existence and role of glassy bridges in determining mechanical response [1,34–36]. For example, work by Robertson et al. on silica-reinforced styrene-butadiene rubber [37] and poly(vinyl acetate) [38] demonstrated that these systems exhibited no alteration of the segmental dynamics of interfacial polymer, while still exhibiting reinforcement relative to the neat matrix. Ultimately, it is difficult to experimentally distinguish the role of the multiple interacting changes induced by filler additives on the elastomer, e.g., physical and chemical polymer-filler interaction, filler dispersion state, and crosslinker distribution. At the least, reinforcement in the absence of suppressed segmental dynamics at the interface [36] suggests that there is more to the mechanism than glassy bridging at relevant temperatures.

More recently, our group has presented evidence for an apparently leading-order mechanism of reinforcement that does not require glassy bridging [39,40]. Specifically, a Poisson's ratio mismatch between the elastomer matrix and nanofiller network in the composite leads to a volume increase on deformation – a consequence of competition between the filler network's tendency to expand on deformation and the elastomer's tendency to conserve volume. Because this volume deviation is largely borne (at a molecular level) by the elastomer matrix, this leads to a contribution from the elastomer's bulk modulus to the elongational modulus of the composite. Because the elastomer's bulk modulus is typically of order of 1000 times its Young's modulus, this contribution can be quite large. This suggests that the leading-order effect of the filler network is to resist *contraction* in the direction *normal* to deformation, with elongational reinforcement emerging via this bulk modulus effect – a scenario closely related to an early direct-particle-contact theory of Witten, Rubinstein, and Colby [17]. Crucially, this mechanism does not rely upon enthalpic interactions between filler particles or between filler and elastomer, because the filler compressive stresses central to this effect are expected to be athermal in nature.

This leaves open a central question: can this mechanism account for the inversion of the temperature dependence of the Young's or storage modulus in the absence of glassy bridges? Here we employ computer simulations and theory to demonstrate that the Poisson's ratio mismatch between the filler network and the elastomeric matrix indeed leads to an inversion of the temperature dependence of the modulus in the absence of glassy bridging. Within this theory, this inversion occurs because the degree of reinforcement is determined by the bulk modulus of the elastomeric matrix. The bulk modulus, in turn, drops on heating due to its enthalpic nature. We demonstrate that this scenario is in quantitative agreement with simulation and predicts the qualitative behavior observed in experiments.

* Corresponding Author E-mail: dssimmons@usf.edu

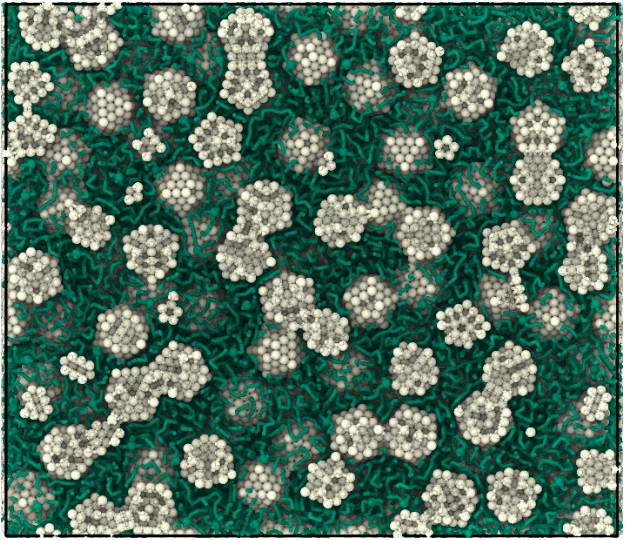

FIG 1. Rendering of a configuration of model filled elastomer with loading of 150 parts per hundred rubber (PHR) or 0.415 filler volume fraction. Polymers are rendered in green (with bonds shown); beads that comprise filler particles are rendered in pale yellow.

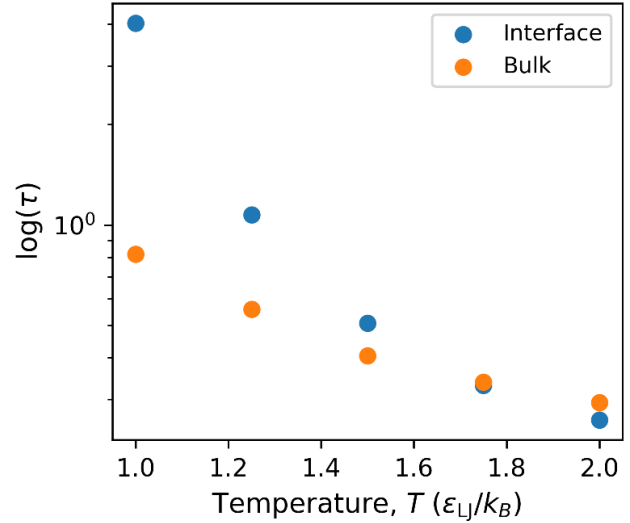

FIG 2. Relaxation times of interfacial and bulk polymer segments (distance less than 1 $\sigma$ and greater than 3 $\sigma$ away from filler surface, respectively) versus temperature in Lennard-Jones (LJ) units ($\varepsilon_{LJ}$ is LJ energy scale and $k_B$ is the Boltzmann constant).

We begin by amplifying the theoretical basis for expecting an inverted temperature dependence of the modulus of elastomers reinforced by compressive particle jamming or networking in the normal direction. Our prior work demonstrated that the modulus of a filler-polymer composite, wherein the polymer is forced to deform at a non-native Poisson's ratio due to normal compressive resistance by the filler network, is given by the following equation [39]:

$$E_c = f\left[E_e + 2K_e\left(\nu_e - \nu_c\right)\right]. \quad (1)$$

Here, $E_c$ is the composite Young's modulus, $E_e$ is the neat Young's modulus, $K_e$ is the neat polymer's bulk modulus, $\nu_e$ is the neat polymer's Poisson's ratio, $\nu_c$ is the composite Poisson's ratio, and $f$ is a strain amplification factor of order 1 arising due to strain localization within any classical composite mechanical theory. In our prior work, this theory was shown to quantitatively predict the reinforcement of the Young's modulus, within the class of simulated filled elastomers employed in the present study [39].

For highly reinforced systems, $E_c$ is dominated by the $K_e$ term. Therefore, because $K_e$ drops on heating, equation (1) predicts that $E_c$ should drop on heating in highly reinforced elastomers, in the absence of glassy bridging effects.

We next report on simulation results demonstrating that this inversion indeed occurs as predicted by this theory. We test this scenario in simulations of neat and filled elastomers, performing molecular dynamics simulations in LAMMPS [41,42]. Our model neat elastomer is made up of 20-bead bead-spring precursor chains, crosslinked at their ends via crosslinker beads, similar to a standard model in the literature [43–49]. To simulate filled elastomers, we introduce to this matrix (before crosslinking) sintered clusters, each comprised of seven primary icosahedral particles, using an approach developed in our prior work to mimic structured carbon black nanoparticles (see FIG 1, rendered in Ovito [50]) [39,40,51,52].

Because our model system is far above its $T_g$ ($T > 2.5$ $T_g$ [53–58]) and has symmetric interactions (i.e., polymer is not selectively attracted to the filler surface), it does not involve glassy bridging, as indicated by studies of $T_g$ effects on local dynamics at high $T$ [37,38]. We confirm this in FIG 2 by computing segmental relaxation times of the polymer matrix near and far from nanoparticles (see End Matter for methodological details). As shown here, the segmental relaxation times differ only by a factor of 5 near the particle and remain much lower than the inverse rate at which we deform our system. Systems containing significant glassy gradient effects typically involve relaxation time gradients that are many orders of magnitude larger [59]. The weak gradients here at $T = 1$ are insufficient to yield substantial glassy bridge effects, as interfacial domains will still readily relax at the simulated deformation rate. Indeed, in our prior work, we showed that reinforcement at a single temperature in systems containing these particles could be quantitatively predicted without reference to the glassy bridging effect, confirming that they are absent [39]. Moreover, this system involves purely repulsive filler-filler interactions. This rules out the presence of appreciable enthalpic attraction effects of any kind between particles. Further information about our initial configurations, forcefield, and simulation protocol can be found in the End Matter and in previous publications [39,40,51].

* Corresponding Author E-mail: dssimmons@usf.edu     

As shown in FIG 3(a), this system exhibits an inversion from stiffening on heating in the pure elastomer to softening on heating in the highly reinforced filled elastomer, as is commonly observed in experiment. This is observed despite the absence of glassy bridging effects, as expected from the Poisson's ratio mismatch theory [39]. As shown by FIG 3(b), the bulk modulus of the neat elastomer indeed drops with heating in this system. To test more quantitatively the origins of the observed inversion in $E_c$, we apply Equation (1) to the data in FIG 3. To do so, we first measure in simulation the neat elastomer Poisson's ratio $v_e$ and composite Poisson's ratio $v_c$ as a function of temperature, via finite difference of extensional strain to normal strain at 5% imposed extensional strain. The resulting values, shown in FIG 3(c), are then combined with the data in the other panels to test Equation (1) against the temperature-dependent data. As shown in FIG 3(d), Equation (1) quantitatively describes the temperature dependence of the modulus. The single (temperature invariant) adjustable parameter in the fit, $f$, is found to have a value of 1.18, which is in excellent accord with values of $f$ measured from chain end-to-end vector deformation for these systems in our prior study [39]. These results indicate that the inversion to softening on heating is indeed quantitatively predicted by the Poisson's ratio mismatch physics and theory described above and is driven by softening of the bulk modulus on heating.

The above results demonstrate that an inverted $E_c(T)$ dependence does not require glassy (or other enthalpic) bridges but instead can emerge from the underlying physics of coexisting particulate and elastomeric solids with distinct preferred Poisson's ratios. With this in mind, we extract from Equation (1) the anticipated temperature dependence of the modulus in highly reinforced elastomers. The temperature dependence of $E_e$ is given by the classical theory of rubber elasticity as $E_e = AT$, where $A$ is proportional to the density of elastically effective strands [28,60]. $K_e$ is expected to drop on heating because it is the inverse of the isothermal compressibility (which grows on heating). This temperature dependence is commonly treated via the Tait equation [61],

$$K_e(T) = \frac{1}{\kappa_T(T)} = \frac{b_0}{C} e^{-b_1 t}, \quad (2)$$

where $\kappa_T$ is the isothermal compressibility, $t$ is the temperature in Celsius, and $b_0$, $C$, and $b_1$ are substance-specific parameters that are tabulated for a wide range of polymers [62]. $v_e$ can be related rigorously to the elastomer bulk and Young's moduli in the low strain limit via a standard Lamé relation:

$$v_e = \frac{1}{2}\left(1 - \frac{E_e}{3K_e}\right). \quad (3)$$

Finally, in the limit of high reinforcement, one expects the dependence of $v_c$ on temperature to become weak due to dominance by the relatively athermal nanogranular network Poisson's ratio. Indeed, as confirmed by FIG 3(c), $\nu_c$ is relatively temperature insensitive for our simulated highly reinforced elastomer. Focusing on this high reinforcement limit ($\nu_c$ is constant) and combining Equations (1) and (3) yields the following equation.

* Corresponding Author E-mail: dssimmons@usf.edu      

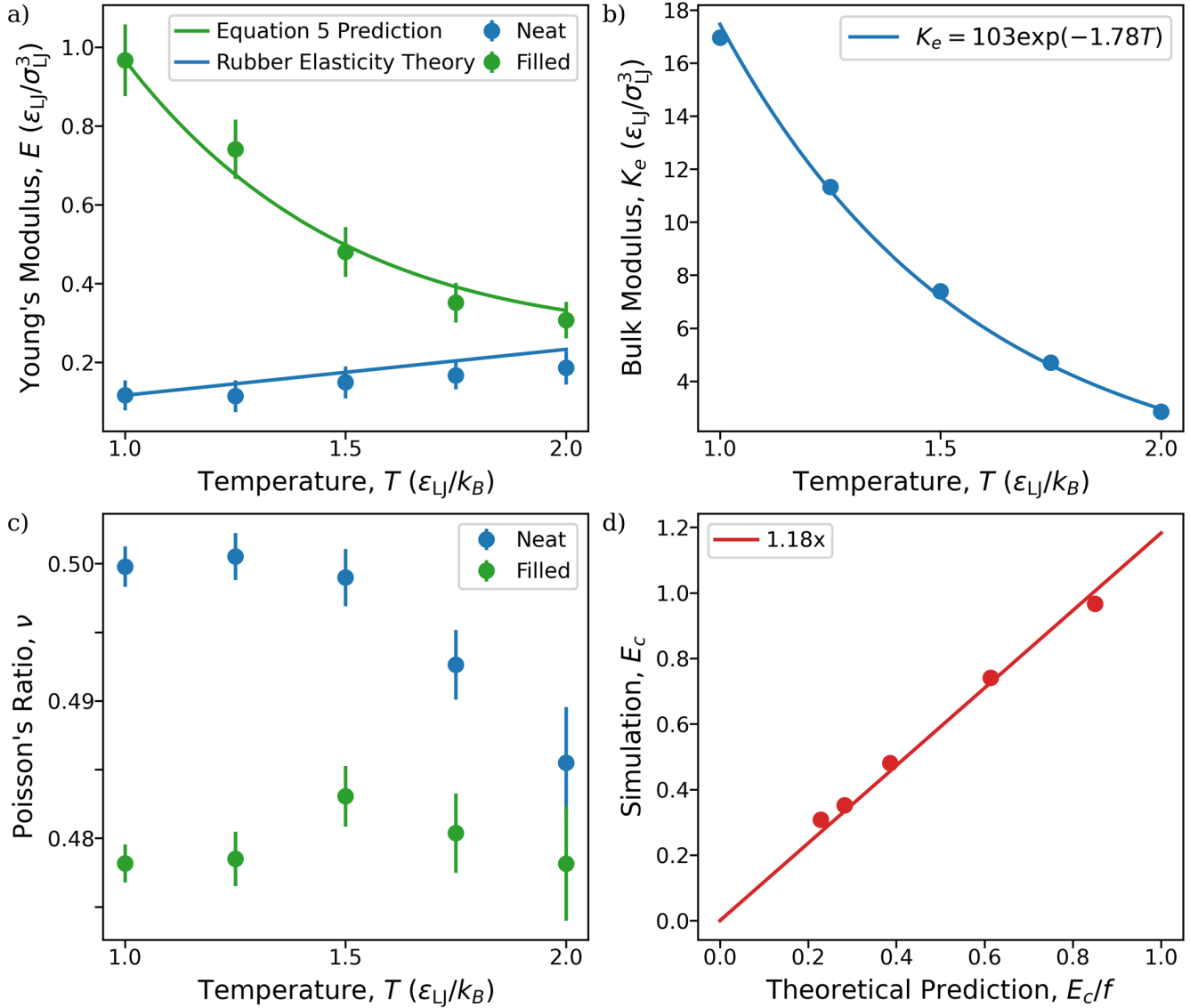

FIG 3. (a) Young's modulus $E$ versus temperature $T$ in LJ units for model neat elastomers and filled elastomers at 150 parts per hundred rubber (PHR) loading or 0.415 filler volume fraction calculated using finite difference of extensional stress between strains of 4.5% and 5.5%. The solid green curve is the prediction of the composite modulus temperature dependence from Equation (5), with $f$ =1.23. The solid blue curve is the prediction of the neat modulus temperature dependence from classical theory of rubber elasticity (i.e., proportionality to temperature) (b) Bulk modulus $K_e$ versus $T$ in LJ units for a model neat elastomer at zero pressure. $K_e$ is computed via the fluctuation-dissipation relation as $T\langle V\rangle/\langle \delta V^2\rangle$, where $V$ is the volume, $\langle V\rangle$ is the NPT ensemble average of the volume, and $\langle \delta V^2\rangle$ is the NPT ensemble variance of the volume. The solid curve is a fit to the Tait equation (Equation (2)). (c) Poisson's ratio $\nu$ versus $T$ computed by finite difference of extensional strain to normal strain between 4.5% and 5.5% extensional strain. d) $E_c$ from panel a) versus theoretical prediction from Equation (1). Error bars reflect standard errors of the mean over 50 (a), 5 (b), and 50 (c) independent replicates.

$$E_c = f\left[\frac{2E_e}{3} + K_e\left(1 - 2\nu_c\right)\right] \quad (4)$$

Furthermore, it is useful to rewrite this equation in terms of the reinforcement at some reference temperature $T_{ref}$. Writing Equation (4) for this reference state, combining it with Equation (4) (written in general for any state as above), and making use of the temperature dependences of $K_e$ and $E_e$ from the Tait equation and classical theory of rubber elasticity, respectively, leads to the following equation:

$$E_c = E_e^{ref}\left[\frac{2T}{3T_{ref}} + \left(\frac{E_c^{ref}}{E_e^{ref}} - \frac{2}{3}f\right)e^{-b_1(T-T_{ref})}\right], \quad (5)$$

where superscripts of "ref" denote the value of the associated quantity at the reference temperature.

* Corresponding Author E-mail: dssimmons@usf.edu

In FIG 3, we test the validity of Equation (5) for our simulated system. The Tait equation is found to provide an excellent description of the temperature dependence of the bulk modulus of the neat system with $b_1$ = 1.78 in reduced LJ units (FIG 3(b)). Taking $T$=1 to be the reference condition, Equation (5) then yields the solid curve in FIG 3(a) as a prediction of the temperature dependence of $E_c(T)$ with $f$ = 1.18 as indicated by our discussion above and consistent with our prior results. This prediction is in excellent agreement with the data.

Experimentally, for diverse elastomers above their $T_g$, $b_1$ is in the range of 0.003 to 0.006 K$^{-1}$ [62]. Consider an elastomer with a 1 MPa modulus at 300 K in its neat state, and a corresponding elastomeric nanocomposite in which introduction of particles enhances this modulus 10-fold at this temperature (i.e. $E_c^{ref}/E_e^{ref} = 10$). Upon heating to a temperature of 400 K, the equation above with Tait equation parameters $b_1$ = 0.0045 K$^{-1}$ predicts a reduction in composite modulus by about 1/3. Over the same temperature range, the neat elastomer modulus would be expected to increase by 1/3 based on classical theory of rubber elasticity. Together, this would predict a reduction to about 5-fold reinforcement relative to the neat elastomer at 400 K. The theory thus predicts a strong heating-induced softening of modulus and relative reinforcement in highly-reinforced polymer composites, of the same close order as the dependence observed in experimental systems [13,22,26,29,30,35].

Our results demonstrate that, rather than requiring the presence of glassy bridges, the inversion of the temperature dependence of the modulus in highly reinforced composites is a natural consequence of the coexistence of a (preferentially volume-conserving) soft elastomer matrix and a (granular like, i.e., $v \cong 1/3$) rigid nanoparticulate network. In essence, reinforcement emerges from a competition between the preferred nonequilibrium volumes of coexisting elastomer and particle networks under deformation. Because the two networks share a boundary condition, they compete over an intermediate compromise volume. This situation, without the introduction of any additional physical postulates, leads to reinforcement that is mediated by the bulk modulus of the elastomer matrix. Whenever this effect dominates the composite modulus, the temperature dependence of the composite stiffness should invert due to natural softening of the bulk modulus on heating, without an explanatory need for glassy bridging or other particle-particle interactions. These theory and simulation results suggest that, rather than pointing to some enthalpic-bridging phenomenon, inversion of $E_c(T)$ is a signature of an underlying mechanism of volume competition between co-continuous soft elastomer and rigid nanoparticulate networks under deformation.

## ACKNOWLEDGEMENTS

This material is based upon work supported by the U.S. Department of Energy, Office of Science, Office of Basic Energy Sciences, under award number DE-SC0022329.

* Corresponding Author E-mail: dssimmons@usf.edu

# END MATTER

*Neat elastomer model*: To study the properties of neat elastomers, we end-crosslink 5000 chains, comprised of 20 Kremer-Grest beads each, via functionality-four crosslinker beads. We crosslink 95% of all possible crosslinks, producing a well-developed elastomeric network [45,46]. Beads interact via a potential model identical to that in our previous works [39,40,51]. Briefly, all polymer beads interact via a 6-12 Lennard-Jones (LJ) nonbonded potential, with interaction energy parameter $\varepsilon_{pp}$ = 1, interaction length scale $\sigma$ = 1, and cut off at a distance $r_{cut}$ = 2.5$\sigma$, yielding attractive interactions. Bonds are modeled via breakable quartic bonds [39,47,48], with parameters tabulated in our prior study [39]. This does not alter findings relative to standard FENE bond model in these simulations, as no bonds break at the low strains probed in this study.

*Filled elastomer model:* We employ model fillers that are clusters of seven icosahedral primary particles. Each of the seven particles contains a center bead with three shells of beads around it to make up 147 beads with 936 breakable quartic bond, as in our prior works [39,40,51]. Filler beads also interact via the 6-12 LJ potential. Filler beads likewise use $\sigma$ = $\varepsilon_{pf}$ = $\varepsilon_{ff}$ = 1 (where $\varepsilon_{pf}$ and $\varepsilon_{ff}$ are the 12-6 Lennard Jones energy parameters for polymer-filler and polymer-polymer interactions, respectively). Filler-polymer interactions are identical to polymer-polymer interactions, with $\sigma$ = $\varepsilon_{pf}$ = 1 and cutoff distance $r_{cut}$ = 2.5 $\sigma$. Filler-filler interactions, however, are fully repulsive, with $\sigma$ = $\varepsilon_{ff}$ = 1 and $r_{cut}$ = $2^{1/6}$ $\sigma$, where this cutoff corresponding to the minimum of the LJ potential. Finally, filler beads are also modeled as breakable quartic bonds, parameterized identically to reference [37].

A single icosahedral particle maps to a diameter of 7-14 nanometers. Primary particles are sintered into clusters of 7 particles by choosing random particle faces for sintering and progressively building from 1 to 7 particles per cluster. The system contains a total of 103 such clusters, each of which is structurally distinct and generated randomly. Anomalously elongated clusters are rejected. This bonding and sintering scheme is identical to that employed in our former work. [39,40] The 103 clusters are dispersed into an elastomer matrix prior to crosslinking to make a filled elastomer with 0.415 filler volume fraction using PACKMOL [52]. An established protocol is used that employs soft nonbonded potentials to avoid overlaps and gradually grow the filler particles into the polymer melt using a combination of NPT and NVT molecular dynamics (MD) simulations. The reader is referred to the Methods section of reference [40] for complete details.

*Simulation methods*: We employ MD simulations performed in LAMMPS to probe the thermodynamic, dynamic, and mechanical properties of model filled elastomers. Simulations employ a time step of $10^{-3}$ $\tau_{LJ}$, where $\tau_{LJ}$ is the LJ time unit (≈1 picosecond). We control temperature and pressure using a Nosé–Hoover thermostat and barostat with damping parameters of $5\tau_{LJ}$ and $2\tau_{LJ}$, respectively. Linear momentum of the center of mass of all atoms is zeroed every 10000 MD steps by adjusting bead velocities to avoid nonphysical drift.

After the neat and filled elastomer configurations are created, they are equilibrated at the temperature of interest and pressure $P$ = 0 for $5.5 \times 10^6$ MD time steps. Finally, 50 thermal replicates are made by saving a configuration every $1 \times 10^5$ MD time steps or $100\tau_{LJ}$ time during an isothermal run. Therefore, each replicate simulation described subsequently begins with a thermally forked configuration. This allows for averaging over stress fluctuations, which are large and short-lived in simulations [63] and prevent calculation of non-noisy low-strain moduli in the absence of a procedure of this kind.

*Analysis methods*: To compute Young's moduli and Poisson's ratios, we measure volume and engineering stress from uniaxial extension deformation simulations with a time step of $10^{-3}$ $\tau_{LJ}$ and an engineering strain rate of $5 \times 10^{-5}$ $\tau_{LJ}^{-1}$, chosen to minimize the rate dependence of the stress and ensure that results reflect rubbery plateau behavior [39,40,51]. During deformation in the uniaxial dimension, the normal pressure is maintained at zero to ensure that the material deforms at its preferred Poisson's ratio. Given the significant fluctuations in stress anticipated in such simulations [63], we sample stress every MD step. In the computation of Young's modulus, we average stress values over 1% intervals strain centered at 4.5% and 5.5% strain and compute the finite difference of engineering stress to engineering strain. In this way, our Young's moduli are tangents computed at 5% strain. Similarly, Poisson's ratios use finite difference of extensional strain to normal strain at 5% extensional strain.

To compute the neat bulk modulus $K_e$, we measure the volume average $\langle V \rangle$ and variance $\langle \delta V^2 \rangle$ from five NPT ensemble simulation replicates of the neat elastomer at each temperature and at zero pressure. Then, we employ the following standard fluctuation-dissipation relationship $K_e = T\langle V \rangle / \langle \delta V^2 \rangle$ [64]. Each replicate is simulated for $2 \times 10^7$ MD steps with a time step of $10^{-3}$ $\tau_{LJ}$.

We compute segmental relaxation times near and far from the filler surface in the following manner. Interface regions are identified as comprising all polymer beads within 1 $\sigma$ of a filler bead (i.e., in the first shell around filler particles), whereas bulk regions are identified as all polymer beads further than 3 $\sigma$ away from the closest filler surface. Relaxation times on the y-axis of FIG 2 are computed using a two-exponential Kohlrausch-Williams-Watts function fit of the self-intermediate scattering function of each region at each temperature, computed at a wavenumber of 7.07 (near the first peak in the segmental structure factor) from an NPT simulation at zero pressure. The relaxation time is then defined as the time at which this fit interpolates to a value of 0.2. This protocol for determining local relaxation times is well-established and exactly matches an approach that has been widely used in studying near-interface gradients in relaxation times and $T_g$ [59,65–67].

* Corresponding Author E-mail: dssimmons@usf.edu

* Corresponding Author E-mail: dssimmons@usf.edu